\documentstyle[12pt,prd,aps]{revtex}
\tightenlines

\newcommand{\be}{\begin{equation}}
\newcommand{\ee}{\end{equation}}
\newcommand{\bea}{\begin{eqnarray}}
\newcommand{\beas}{\begin{eqnarray*}}
\newcommand{\eea}{\end{eqnarray}}
\newcommand{\eeas}{\end{eqnarray*}} 
\newcommand{\ba}{\begin{array}}
\newcommand{\ea}{\end{array}}
\def\ls{\mathrel{\lower4pt\vbox{\lineskip=0pt\baselineskip=0pt
           \hbox{$<$}\hbox{$\sim$}}}}
\def\gs{\mathrel{\lower4pt\vbox{\lineskip=0pt\baselineskip=0pt
           \hbox{$>$}\hbox{$\sim$}}}}

\begin{document}

\draft



\title{Electric Charge Quantization and Extra Dimensions}

\author{
A. P\'erez-Lorenzana$^{1,2}$\footnote{e-mail:aplorenz@ictp.trieste.it} 
and C. A. de S. Pires$^{3}$\footnote{e-mail:cpires@ift.unesp.br}}

\address{
 $^1$ The Abdus Salam International Centre for Theoretical Physics,
I-34100, Trieste, Italy\\
$^2$  Departamento de F\'\i sica,
Centro de Investigaci\'on y de Estudios Avanzados del I.P.N.\\
Apdo. Post. 14-740, 07000, M\'exico, D.F., M\'exico\\ 
$^3$ Departamento de F\'{\i}sica, 
Universidade Federal da Para\'{\i}ba, \\
Caixa Postal 5008, 58051-970 Jo\~ao Pessoa PB, Brazil.
}

\date{May, 2002}

\maketitle

\begin{abstract}
In models with flat extra dimensions tiny Dirac neutrino masses can be
generated via the coupling of four dimensional 
Standard Model fields to a higher dimensional fermion.
Here we argue that, in spite of the Dirac nature of the neutrino, 
quantization of the electric charge can still be understood  
as a result of  anomaly cancellation, charge
conservation and naturalness requirements.
\end{abstract}
\pacs{14.60.Pq; 14.60.St;11.10.Kk}


\section{Introduction}

One of the most intriguing mysteries of nature is the empirical
observation that the electric charge of the fundamental particles appears
to be quantized. This question has already attracted some attention in the
past and some possible ways to understand it are already known. The first
suggestion was given by Klein~\cite{klein} at the beginning of the past
century through the introduction of extra dimensions. Another proposal
came from Dirac~\cite{dirac} who linked electric and magnetic monopole
charges. A third possibility arises in the context of Grand Unification
Theories~\cite{guts} where the quantization of charge comes naturally as a
plus from the group structure.

More recently, in the context of four dimensional theories, 
some further attempts to understand the basics of this
problem have addressed the question in a more general
framework~\cite{foot1,moha1,moha2}:  In a large class of models that
include a U(1) gauge group factor contributing to electric charge, its
quantization is not always guaranteed. The problem is that the
U(1) generator can a
priori take a continuous set of values. However, classical as well as
quantum constraints may restrict such values. Namely, the requirement of
having massive fermions and anomaly cancellation may, for instance,
explain the values of the hypercharge taken by matter fields in a single
family in the Standard Model (SM). Though one may take this as a
cheerful notice, two facts should be yet considered.  First, in the SM
neutrinos are massless, which alone seems not to be in good agreement with
the  observational evidence of neutrino oscillations in
solar~\cite{sol}, atmospheric~\cite{atm} and terrestrial~\cite{lsnd}
neutrino experiments. Second, unless one makes the extra assumption that all
families are exact replicas of each other, in the sense that similar
representations have equal hypercharge, the electric charge quantization
in the actual SM with three families seems again ad hoc. In fact, in this
case there exist some hidden symmetries which, being anomaly free,
insert an arbitrariness in the definition of the electric 
charge~\cite{foot1,moha1}.
In order to restore electric charge quantization one has to
break those hidden symmetries. The simplest way to do
it is by given the neutrino a Majorana mass~\cite{moha1}. Indeed, if
neutrinos are of a Dirac type the electric charge appears  to be 
dequantized. This is due to the
introduction of  extra degrees of freedom, the right handed neutrinos,
which come with out adding extra constraints to the theory.

In this short note we are interested in analyzing these results 
in the context of theories that involve large 
extra dimensions~\cite{nima1}. We are mainly motivated by the fact
that Dirac neutrinos are a very likely byproduct of such 
theories~\cite{nima2,3nus,others,langa}.
Our main observation here is as follows: Even though at first sight a
Dirac neutrino does not seem to go well with electric  charge
quantization,  in  models with extra dimensions the requirement of charge
conservation and  naturalness on the explanation to neutrino masses
introduce an extra constraint on the assignment of hypercharges  
that leads
to the restoration of  electric charge quantization. 
Shortly, they  fix 
the hypercharge of the right handed neutrinos to be null. 
This is a dynamical version of the same condition introduced when one
writes (by hand)  Majorana mass terms for the neutrinos.
To make our point, and for completeness, 
we shall first review  the  electric charge quantization  in the 
usual four dimensional SM with
one family, and  explicitly show how the introduction of a right handed
neutrino state with only a Dirac coupling dequantize the electric charge. 
Next, we shall analyze
the case of models with extra dimensions and argue that here, even if
neutrinos are Dirac fields, the  electric charge quantization is 
restored, provided that the right handed neutrino is a bulk field.
Further remarks on these theories are added by the end.

\section{Charge quantization and neutrino mass}

Let us start by reviewing the well known case of the 
SM with a single family.  Due to the
abelian nature of hypercharge, $Y$, 
the assignment of charges to matter fields is in
principle arbitrary. However, symmetry breaking triggered by the
acquisition of a vev by the  Higgs
doublet, $H$, relates $Y$ with electric charge
through the formula
 \be 
  Q = T_{3L} + {1\over 2}{Y\over Y_H}~;
  \ee
where $Y_H$ is the $H$ hypercharge, which can always be normalized to
$Y_H=1$. For the fermion sector we take
 \bea
  Q_L: \quad (3,2,Y_q), &\qquad& u_R: \quad (3,1,Y_u), \nonumber \\
  d_R: \quad (3,1,Y_d), &\qquad&   L : \quad (1,2,Y_\ell), \\
  e_R: \quad (1,1,Y_e) ~.  \nonumber
 \eea
Now, the requirement of renormalizability of the gauge theory, as well as
the  existence of Higgs couplings that generate fermion masses, introduce
a number of constraints among the above hypercharges that further reduce
their arbitrariness. We first consider the Yukawa coupling terms
\[ 
{\cal L}_Y = h^\ell \bar L H e_R + 
              h^u \bar Q_L \tilde H u_R + 
              h^d \bar Q_L H d_R + h.c.
\]
They give
 \be
 \label{const1}
  Y_e = Y_{\ell} -1 ; \quad  Y_u = 1 + Y_q; \quad \mbox{ and }\quad 
  Y_u = Y_q -1 ~.
 \ee
Cancellation of the triangular quantum anomalies, 
name\-ly 
$Tr~U(1)_Y[SU(3)_c]^2$; $Tr~U(1)_Y [SU(2)_L]^2$; $Tr~[U(1)_Y]^3$;
and $Tr U(1)~\times$~{\it grav.~anomaly};
contribute with only two independent constraints, 
which combined with those above can be written as
 \bea
Tr~U(1)_Y [SU(2)_L]^2&\Rightarrow& 3 Y_q + Y_\ell =0 ;\nonumber \\
 \mbox{ and }\quad  Tr~[U(1)_Y]^3&\Rightarrow& Y_\ell = -1~.
 \label{anom1}
 \eea
Other anomalies are
identically canceled with the help of the above conditions.
Notice that there are as many constraints as free parameters in our
analysis.  Therefore, a unique solution exist which gives  the well
known values: 
\bea 
Y_q=1/3~;&\quad& Y_u = 4/3~; \quad Y_d= -2/3~; 
\nonumber\\
Y_\ell =-1~; 
&\quad& Y_e = -2~.
\label{ys}
\eea
Hence, this procedure provides a  natural explanation to quantization of
charge.  Nevertheless, as already pointed out, 
this theory contains only  massless neutrinos.

\subsection{Dequantization by a Dirac neutrino}

By minimally extending the spectrum of the theory to contain a right
handed  weak and color fermion singlet, $\nu_R:~(1,1,Y_\nu)$, we  allow
for
a new Yukawa coupling of the form $LH\nu_R$. Thus, besides the constraints
in Eq.~(\ref{const1}) we now also have
 \be 
 \label{const2}
  Y_\nu = 1 + Y_\ell~.
  \ee
However, now  anomaly cancellation  yields to  only one
extra constraint on $Y_\ell$ and $Y_q$, 
the one associated to  $Tr~U(1)_Y [SU(2)_L]^2$: $3 Y_q + Y_\ell =0$. 
All other anomaly expressions  become just identities.
Therefore, such theory still has one free parameter, $Y_\ell$, 
which can be fixed arbitrarily. This 
spoils the quantization of the  electric charge. 
This phenomenon is usually referred as charge dequantization and  it is
associated to the presence of some hidden (anomaly free) 
global symmetry in the theory ~\cite{foot1,moha2}, here 
identified as  $U(1)_{B-L}$.

If the theory is assumed to be an effective theory in which a 
Majorana mass term, $\bar\nu_R^c\nu_R$, is present, this adds the lacking
constraint, fixing the extra degree of freedom, by explicitly breaking 
the $U(1)_{B-L}$ symmetry. Clearly, writing this term is equivalent
to take $Y_\nu =0$. This 
fixes $Y_\ell = -1$, and other charges follow as needed. 

\subsection{Dequantization  with three families}

The above analysis can be straightforwardly generalized for three families.
Notice, however,  that by writing the Yukawa couplings one should
keep in mind that there is a mixing in the quark sector, 
parametrized by the CKM matrix. Thus, one gets
 \[
 {\cal L}_Y = h^\ell_{ii} \bar L_i H e_{iR} + 
              h^u_{ij} \bar Q_{iL} \tilde H u_{jR} + 
              h^d_{ij} \bar Q_{iL} H d_{jR} + h.c.
 \]
From here, the hypercharges of the quark fields should obey
$Y_{ui}= Y_u$; 
$Y_{di}=Y_d$ and 
$Y_{qi}=Y_q$. 
This reduces the number of constraints to
$Y_u = Y_q +1$; $Y_d = Y_q -1$  and  
$Y_{e i} = Y_{\ell i}-1$;
for the family indices $i,j=1,2,3$. 
Once more, requiring  cancellation of
anomalies is not enough as to uniquely define the hypercharges. 
As before, only $Tr~U(1)_Y [SU(2)_L]^2$; and  $Tr~[U(1)_Y]^3$ 
give non trivial conditions, which added to those above may only fix seven
of the remaining nine free parameters. 
Again,  the charge is not quantized.  
The hidden symmetries are those associated to the lepton number
combinations: $L_e - L_\mu$; $L_e - L_\tau$; $L_\mu - L_\tau$. The simplest
way to break these symmetries is allowing for a general mixing in the
leptonic sector as we did for the quark sector, since their presence adds
extra constraints, namely $Y_{ei}= Y_e$, and $Y_{\ell j}=Y_{\ell}$. 
Thus, reducing the problem to the single family case.  Such
mixings appear naturally if neutrinos are massive, but again, 
if neutrinos
are Dirac-like, $B-L$ will reappear as a hidden symmetry that, once
more, plays against charge quantization.

\section{Bulk neutrino and charge quantization}
\subsection{Neutrino mass in extra dimensions}

All above observations seem not to go well with the possibility that the
neutrinos be Dirac-like particles. At first glance this is a  good
motivation to assume that they are rather Majorana particles.
A second motivation for a Majorana neutrino comes from the fact that 
a Dirac mass needs a large  fine tuning to provide eV masses, 
whereas Majorana masses  can be explained  quite naturally  by the
see-saw mechanism~\cite{seesaw}. 
So, most of the current models for explaining  neutrino masses and  
mixings  follow this trend.

Nevertheless, in theories with
extra compact dimensions  the situation is on the opposite.  In those
models  SM particles are assume to live on a four dimensional hypersurface
(the brane) embedded in a higher dimensional space (the bulk). The extra
dimensions are here taken to   be compactified on a
flat  manifold~\cite{nima1}. These theories have been motivated by the
possibility of having a small fundamental scale for quantum gravity.
The aftermath of such 
constructions is the reduction of the energy scale cut-off that suppresses
all  non renormalizable operators that involve SM particles, as for instance  
the dimension 5 operator  that gives a Majorana mass to the neutrino:
 \be 
 \label{op1}
 {(LH)^2\over \Lambda}~.
 \ee
The physical meaning of the scale $\Lambda$ depends 
on the nature of the compactification as well as on 
the physics that generates  such an operator.
In theories with flat extra dimensions, $\Lambda\leq M$, where $M$ is the
fundamental scale at which gravity becomes strong, 
that  related to the effective Planck scale $M_P$ and the volume of the 
n-th dimensional extra compact space, $V_n$, 
by the relationship~\cite{nima1}: $M^{n+2} V_n = M_P^2$. 
Current limits indicate that $M$ could be  as low as 
few TeV~\cite{exp1,rizo}.

With such a small $\Lambda$ the neutrino mass generated by
the operator in Eq.~(\ref{op1}) comes out to be too large.  
Thus, such operators have to be avoided, say, by
imposing lepton number conservation. 
Instead, naturally small Dirac masses may be generated by introducing a bulk
neutrino~\cite{nima2}. 
Now, despite of having only Dirac neutrinos, 
our previous understanding of the 
electric charge quantization can remain 
due to the own characteristics of these models. 

To illustrate our claim let us 
briefly mention how light Dirac neutrinos are introduced in these
theories.
As a consequence of the localization of the SM fields on the brane, all
their couplings to fields that freely propagate in the whole space  get a
volume suppression, so that they  become very small in the effective four
dimensional theory. This is the reasoning  applied for understanding  the
smallness of neutrino mass. Let us consider  a right handed bulk neutrino,
$\nu_{BR}$, that couples to the SM fields through the Yukawa coupling
 \[ 
 \int\!d^4x\, d^ny  {\tilde h\over \sqrt{M^n}}~\bar L(x) \tilde 
   H(x) \nu_{BR}(x,y)\, \delta^n(y)~,
 \]
with  $\tilde h$ the Yukawa coupling. There we 
have assumed that the SM brane is localized at the position
$y=0$, where $y$ represents the $n$ extra space coordinates. 
This coupling explicitly conserves total Lepton number and gives the same
constraint for the hypercharges as in Eq. (\ref{const2}).
The zero mode wave function  of the bulk neutrino field goes as
$\nu_{0BR}(x)/V_n^{1/2}$. Then, 
after integrating out the
extra dimensions and  setting in the Higgs vev, $v$, one gets 
a small Dirac mass given as 
 \be m_D = \tilde h {M v \over M_P}~. \ee
Notice that  for $M\sim 1$~TeV one gets  a
mass of order $10^{-5}$ eV~\cite{nima1} for $\tilde h =1$. 
A larger fundamental scale, however,  or moderately strong bulk coupling 
would produce the right order of masses for explaining neutrino anomalies.
Moreover, 
requiring that $m_D\ls .1$~eV to account for the range of mass parameters
needed to fit the experiment, one arises to the condition that 
 \be  \label{h} \tilde h = { M_P\, m_D \over  M\, v}
 \ls 10^{-12} \left({M_P\over M}\right)~. \ee

\subsection{Naturalness and charge quantization}

There is a crucial ingredient in the scenario we are considering:
SM gauge interactions are attached to the brane. 
In such an case any bulk field should be neutral under the SM groups, 
including the hypercharge. This  is  to insure  the gauge invariance of the
theory. In other words,  if we allow a hypercharged field to propagate in the
bulk,  by assuming that our right handed neutrino has a non zero $Y_\nu$ for
instance,  then, we would be  forced to promote the hypercharge to be a bulk
symmetry.
Here, a comment is in order. With $U(1)_y$ in the bulk, 
the radius of the extra
space can not be larger than few hundred $GeV^{-1}$, which   makes the
fundamental scale  much larger, 
say than $10^7$~GeV for $n=6$ to $10^13$~GeV for $n=1$. 
With such a large scale one may doubt about the need of the bulk neutrino, 
since the suppression on the operator in Eq.~(\ref{op1}) is now much less sever.
It is interesting, however, to analyze whether even under this circumstances the
bulk neutrino hypothesis can still be at work to provide light Dirac neutrinos.
This program immediately  fails due to the 
volume suppression in  bulk-brane couplings, which implies that the bulk theory
should be  strongly coupled.
To clarify this point let us  consider
the   coupling of the $U(1)_Y$
gauge field, $B_\mu(x,y)$, to the lepton current 
$j_\mu(x) = \bar L\gamma_\mu L$, which  has the form
 \be 
  {g_5\over \sqrt{M^n}} ~  Y_\ell~B_\mu(x,y) j^\mu(x) \delta^n(y)~.
 \ee
In the above formula the suppression comes due to the larger mass
dimensionality  of the  Gauge field: 
$[B] = 1 + n/2$, which is reflected on the
scaling of the  Kaluza Klein modes. At the zero mode level one gets 
$B_\mu(x,y) = B^0_\mu(x)/ \sqrt{V_n}~+$ (higher levels). Therefore, after
dimensional reduction one obtains an effective four dimensional 
theory where the coupling
constant of the hypercharge is given by 
 \be \label{g}
 g = \left({M\over M_P}\right) g_5;
 \ee
As one knows that $g$ is close to one, it comes that  
$g_5$ has to be large enough
as to absorb the suppression on the right hand side.  

Solving the neutrino mass problem with bulk neutrinos, however,  needs an 
unnaturally  large hierarchy between $g_5$ and $\tilde h$.  
In fact, by looking at Eqs.~(\ref{h}) and (\ref{g}) one gets
 \be 
 {\tilde h \over g_5}\sim {m_D\over gv}\ls 10^{-12};
 \ee
which is the 
same hierarchy  one wished to explain since the beginning,
independently of the actual value of the fundamental scale. 
In other words, promoting the
hypercharge  to be a bulk interaction jeopardizes our
former explanation of the neutrino masses with a bulk neutrino.
Hence, the consistency  of the theory 
requires that the hypercharge should be confined to the brane 
as well as the other  SM gauge interactions. 
Moreover, as the right handed  neutrino would still come from the bulk, it
would be forced to be totally chargeless under any SM group 
in order to keep gauge invariance.  

As it is clear,  above arguments mean that $Y_\nu=0$. This adds the
missing constraint to the system of Eqs.~(\ref{const1})  and
(\ref{const2}), obtained from Yukawa couplings and  anomaly cancellation
conditions. 
Straightforwardly one gets, from Eq.~(\ref{const2}), that $Y_\ell
= -1$, which  gives the right values of $Y$ as  in
Eq.~(\ref{ys}). Therefore, our understanding of quantization of charge would
remain though the neutrino is Dirac particle.
Considering three families will give us the same answer.

\section{Outlook}

\subsection{neutrino puzzles}

In connection with the possible explanation of neutrino anomalies,
we would like to add few comments here. 
First, our whole discussion is only
addressing  the problem of preserving  electric 
charge quantization given the presence
of only Dirac mass terms on these theories. 
A simple and 
consistent explanation of both, solar and
atmospheric neutrino data  is actually 
possible in the present context~\cite{3nus,others}, though
the theory will require the existence 
of at least two bulk neutrinos to accommodate the two required 
mass parameters.
However, an explanation to 
LSND results is excluded~\cite{3nus}, 
at least in the minimal version of  these models
(with three right handed neutrinos),
since only two independent
squared mass differences
can be  produced out of three mass eigenstates. 
Nevertheless, there is still the possibility that a fourth bulk neutrino
exist that could provide the extra degree of freedom to accommodate LSND, 
as already suggested in Ref.~\cite{langa}
Notice that this conclusion is regardless 
the actual size of the common radius $R$ in
theories with large extra dimensions~\cite{3nus,others}, 
mainly because Dirac mass terms only depend on 
the ratio  $M/M_P$.
Another consequence of these scenario would be the non observation of
neutrinoless double beta decay. 

\subsection{An anomalous B-L?}

{}As a final note, we observe that $B-L$ 
does not
trivially arise as a gaugeable (anomaly free) symmetry any more. 
The clear example is the
(simplest) case with one large extra dimension compactified on $S^1$.  
There, the bulk theory is vector like.
For higher dimensions one has to cancel all anomalies in the bulk to
keep  gauge invariance. This usually needs the introduction of new extra 
fields. In the simplest case we would  just assume that the bulk theory is
vector-like, independently of the number of extra dimensions.
Thus, there would no contribution to $B-L$ anomaly  coming from
bulk fields. 
However,  with only the SM particle content,  $B-L$ is
anomalous. Indeed, for SM matter fields one has Tr~$U(1)_{B-L}^3 = -3$.
Such anomaly appears localized on the fixed point where the SM
lives. 

The naive approach of introducing orbifolds to project
out chiral zero mode components of the bulk neutrino  
does not seem to help in canceling the $B-L$ anomaly. 
Indeed, in five dimensions, for instance,
such theory develops localized anomalies that sit on 
the fixed points of the orbifold. 
Such anomalies, however, do
not compensate the one developed
by the SM fields~\cite{anomalies}, making the whole theory non gauge
invariant under $B-L$. 
In fact, with a $Z_2$ orbifolding of the circle under
which $y\rightarrow -y$,
the bulk neutrino transforms as $\nu_B\rightarrow \gamma_5 \nu_B$. Thus, at the
fixed point $y=0$, $\nu_{BL}$ vanishes. Nevertheless, it also thus at the other
fixed point at the end of the space, located at $y=\pi$. This accidental 
symmetry of the bulk $\nu_{BL}$ field is reflected in the 5D anomaly, which is
actually localized at the orbifold fixed points:
 \[
 \partial_MJ^M(x,y) = {1\over 2 } \left [ \delta(y) + \delta(y-\pi)\right] 
 {\cal A}(x,y)~,
 \] 
where $J^M$ is the 5D current and  ${\cal A}\sim F\cdot\tilde F$ 
is the 4D chiral anomaly. 
The SM anomaly contribution, on the other hand, would be 
 \[ \partial_MJ^M(x,y) = -\delta(y)  {\cal A}(x,y)~, \]
which obviously does not compensate the 5D contribution.
There is, of course the possibility of removing
such an anomaly by increasing the number of bulk fields. However this may
also affect our above understanding of charge quantization by the
introduction of new degrees of freedom. 
A further analysis of this problem may deserve a further study.


\section{Concluding Remarks}

Along this short note we have argued that, in the context of models with
large compact extra dimensions, Dirac-like neutrinos are consistent with our
understanding of electric charge quantization from gauge anomaly
cancellation arguments. The reason is twofold: First, naturalness
argument on the smallness of the  neutrino mass, 
as generated via bulk-brane couplings, 
force all SM gauge interactions to be fixed to the brane.
Second, in order to get a consistent gauge invariant theory, bulk fermions
must be neutral under all SM groups, hypercharge included. This fixes the
hypercharge of the bulk  neutrino to be null. The
straightforward output of this constraint is the restoration of the
electric charge quantization.  
However, we should notice that on these theories $B-L$ does not seem to be
an anomaly free symmetry, and thus, it can not be consistently gauged, at
least within the minimal context we have considered along our discussion.
This mark a clear difference with the four dimensional models. Further
addition of other higher dimensional terms/fields to the action may be
needed to insure anomaly cancellation. That would be the case of two extra
dimensions, for instance, where antisymmetric tensor fields have to be
added. 

\vskip1em

{\it Acknowledgements.}   
We would like to thank R.N. Mohapatra for comments.
APL also  wishes to thank R.N. Mohapatra 
and   the University of Maryland 
particle theory group for 
the warm  hospitality and support 
during the last stages of this work. 
The work of CP is supported by Funda\c c\~ao de
Amparo \`a Pesquisa do Estado de S\~ao Paulo (FAPESP).


\end{document}